\documentclass[twocolumn]{aa}
\usepackage{graphicx}
\usepackage{psfig}
\usepackage{natbib}
\usepackage{amsfonts}
\usepackage{amssymb}
\usepackage{amsmath}
\usepackage{indentfirst}

\bibpunct{(}{)}{;}{a}{}{,}
\newcommand{\ud}{\mathrm{d}}

\begin{document}

   \title{Which acceleration process for Ultra High Energy Cosmic Rays in Gamma Ray Bursts ?}

\titlerunning{Which acceleration process for UHECRs in GRBs ?}

   \author{D. Gialis
          \inst{1}
          \and
          G. Pelletier\inst{1,2}
          }

   \offprints{D. Gialis}

   \institute{Laboratoire d'Astrophysique de Grenoble\\
              414 rue de la Piscine, Domaine Universitaire, BP 53, F-38041 GRENOBLE CEDEX 9\\
              \email{Denis.Gialis@obs.ujf-grenoble.fr}
         \and
            Institut Universitaire de France\\
             \email{Guy.Pelletier@obs.ujf-grenoble.fr}
            }

   \date{Received ; accepted }

   \abstract{
   In this paper, we have made an accurate investigation of proton acceleration in GRBs and we have predicted a possible signature of cosmic rays, in a sufficiently baryon-loaded fireball, via GeV $\gamma$-ray emission produced by $\pi^{0}$-meson decay. If two ungrounded assumptions are removed, namely, Bohm's scaling and a slow magnetic field decrease, the usual Fermi processes are unable to generate ultra high energy cosmic rays (UHECRs) in GRBs. We propose to develop another scenario of relativistic Fermi acceleration in the internal shock stage. We present the results of a realistic Monte-Carlo simulation of a multi-front acceleration which clearly shows the possible generation of UHECR. The amount of energy converted into UHECRs turns out to be a sizeable fraction of the magnetic energy.
   }

   \maketitle
%

\section{Introduction}     
     
Gamma Ray Bursts (hereafter GRBs) have been considered as promising sources of Ultra High Energy Cosmic Rays (UHECRs) either through the external ultrarelativistic shock \citep{Vietri95} or through the internal shocks \citep{Waxman95}. The acceleration at the external shock does not achieve this goal when the external medium is a standard interstellar medium \citep{Gallant1999}. As for the internal shocks, first and second order Fermi processes would achieve this goal with the extreme assumption of a cosmic ray mean free path of the order of its Larmor radius (so-called Bohm's scaling). In a previous paper \citep{GialisPelletier03}, we have shown that such scaling is physically inconsistent for two reasons; first, the law is not validated by numerical simulations \citep{Casse01} and second, even if inappropriately used, it would lead to tremendously fast magnetic energy depletion in the fireball.     
 We also have considered a more realistic Kolmogorov scaling but a severe expansion loss limitation does not make it possible. These arguments are developed in Sects. \ref{sec2} and \ref{sec3} of this paper, together with the important issue of the magnetic field profile. Because both the global energy budget of the fireball and the Hillas criterium are in favour of the generation of UHECRs, we have made a detailed investigation of a different Fermi process proposed by \cite{PelletierK00} and \cite{Pelletier99}. In this kind of Fermi process, reviewed in Sect. \ref{sec4}, the sheets invoked in the internal shock model are considered as relativistic hydromagnetic fronts that scatter cosmic rays. Thus, cosmic rays undergo a relativistic Fermi acceleration, similar to a kind of second order process through multiple interactions with all the fronts in the relativistic wind. Actually, this is not a second order Fermi process because no expansion is allowed in the relativistic regime, the energy jump being large at each scattering. The generation of cosmic rays resulting from these multiple scatterings is very different from generation obtained by a superposition of the individual first order contributions of each shock.  \\     
In Sect. \ref{sec5} we describe a realistic Monte-Carlo simulation of this process, taking into account several dynamical parameters in a conical expanding fireball, for several intensities and profiles of the average magnetic field. The results are presented in Sect. \ref{sec6}; they suggest a good efficiency of the process of acceleration of UHECRs and give a prediction of the time at which they are created during the internal shock stage. Finally, we establish a possible diagnosis of cosmic rays in GRBs through $\gamma$-ray emission by $\pi^{0}$-meson decay which would be easily detected by HESS observatory, 5@5 experiment and GLAST.

\section{Preliminary considerations}     
\label{sec2}     
     
The goal of the generation oif UHECRs in GRBs is to reach a maximum energy close to the confinement energy limit above which the Larmor radius becomes larger than the correlation length of the magnetic field, namely      
\begin{eqnarray}     
\epsilon_{cl}=Z\,e\,\Gamma\,B\,\ell_{c}\,.   
\label{ecl}     
\end{eqnarray}    
In Eq. (\ref{ecl}), $B$ is measured in the co-moving frame; the correlation length, $\ell_{c}$, also measured in the co-moving frame, stretches with the expansion of the fireball. This length keeps the same order of magnitude as the characteristic size of the subshells and follows the same scaling, where the Lorentz factor is related to the velocity $\beta_0$ through $\Gamma =1/\sqrt{1-\beta_0^{2}}$. This is the reason why we assume $\ell_{c}\sim \Delta R \sim r/\Gamma$ taking $r$ as the distance of a fluid element from the explosion center in the observer frame; $\Gamma$ is the bulk Lorentz factor after the saturation radius.  Hereafter, we suppose that $\Gamma$ reaches the maximum value $\eta=E/M_{b}\,c^2$ where $E$ is the total energy emitted and $M_{b}$ is the total baryonic mass ejected.      
Then, Equ. (\ref{ecl}) becomes     
\begin{eqnarray}     
\epsilon_{cl}\simeq 10^{21}\,\left( \frac{B}{10^{6}G}\right)\left(\frac{r}{10^{-6}pc}  \right)\, eV\,.     
\label{ecl1}     
\end{eqnarray}     
The Fermi acceleration by internal shocks starts at a distance of the order of $r_{b} \equiv \eta^2 r_0$ (about $10^{12}$ cm) which is currently defined as the broadening radius i.e the distance after which the shells broaden. Thus, in this model, a magnetic field of at least $10^{6}$ G is necessary at this radius to achieve UHECR production. Since the product $B\,r$ decreases like $r^{1-\alpha}$ (with $1\leq \alpha \leq 2$), super-GZK protons could be produced between $r_{b}$ and 100 $r_{b}$ for $\alpha = 3/2$ and between $r_{b}$ and 10 $r_{b}$ for $\alpha = 2$. We note that no generation of UHECRs is possible for $B$ smaller than a few $10^{5}$ G in the internal shock model. Regarding the Fermi acceleration at the external ultrarelativistic shock, its intensity  must reach 100 G at the deceleration radius ($\sim 10^{-2}$ pc) corresponding to the equipartition value with the pressure of the shocked plasma. Let us examine this case first.

\subsection{External shock}     
     
The Fermi acceleration of supra-thermal particles at the external ultrarelativistic shock was suggested by \cite{Vietri95} and it has  recently been reviewed by \cite{Vietri2003}. As mentioned before, a 100 G magnetic field is required to obtain UHECRs. If the usual interstellar medium is considered, with a $\mu$G magnetic field, even with the shortening of the acceleration time due to the ultrarelativistic shockfront, namely $t_u \sim t_L/\Gamma_s$, as shown by Gallant \& Achterberg (1999), UHECRs cannot be produced. Moreover the amplification of the  pre-existing cosmic ray distribution does not provide enough UHECR flux. The only hope is through the consideration of a different external medium, as for instance a pulsar wind bubble such as considered by \cite{Gallant1999}.     
     
\subsection{Internal shocks}     
\label{2.2}    
     
The internal shock model has been designed  \citep{Meszaros93} to account for variability observed in the light curve. Indeed millisecond variability, as observed in some GRB light curves, cannot be explained in the framework of a single sheet colliding with the interstellar medium. The most interesting feature of the multiple sheet model is that the time variability at the origin of the flow is observable thanks to the following scheme. Let us consider two sheets that leave the central engine separated by a time interval $\Delta t$, respectively with Lorentz factor $\Gamma_{1}$ and $\Gamma_{2}$. A collision occurs when $\Gamma_{2}>\Gamma_{1}$ at a timr that can be expressed by     
\begin{eqnarray*}     
t_{c}\simeq \frac{2\Gamma_{1}^{2}\,\Gamma_{2}^{2}}{\Gamma_{2}^{2}-\Gamma_{1}^{2}}\, \Delta t \, .     
\end{eqnarray*}     
Assuming an instantaneous shock pulse, some time spreading, $\Delta t_s=t_c/2\Gamma^2$, is observed and thus this duration corresponds to the original time interval.     
The shortest variabilities ($\sim$ 1 ms) correspond to the typical time scale associated with the size of a black hole of a few tens of solar masses (namely $r_{0}/c$). Such a first collision takes place around the distance $r_{b}$. Longer variations correspond to collisions at a more remote distance up to a maximum distance determined by the duration of the flow $\Delta t_w$. This maximum distance is $r_{max} \sim r_{b}\, c\, \Delta t_{w}/r_{0}$, with $c\, \Delta t_{w}/r_{0} \simeq 3\times 10^3(\Delta t_w/1s)$, which gives the proper length of the flow in the co-moving frame $\ell_{0}=\beta\,c\,\Gamma\,\Delta t_{w}$. The duration of the flow during the internal shock phase is therefore $\Delta t_{max}\sim (r_{b}/r_{0})\,\Delta t_{w} \sim \eta^{2}\,\Delta t_{w}$. This phenomenon is observed during a time interval shortened by the propagation effect, namely $\Delta t_{obs}=(1-\beta)\, \Delta t_{max} \simeq \Delta t_{max}/2\eta^{2} \sim \Delta t_{w}$. The previous value of $r_{max}$ determines the range of the acceleration process that extends from $r_b$ to $r_{max}$ which is not far from the deceleration radius $r_d \sim 10^{16}$cm. \\       
     
The Fermi acceleration (first or second order) in the internal shock model is usually considered \citep{Waxman95} as mildly or sub-relativistic with a characteristic time proportional to the Larmor radius (Bohm scaling). However, in a recent paper \citep{GialisPelletier03}, we have shown that this assumption is not realistic regarding the magnetic energy depletion time. Moreover, the Fermi acceleration time depends on the mean free path, $\bar{\ell}$, of the particle in an irregular magnetic field. This length depends on two other lengths, namely the Larmor radius, $r_{L}$, and the correlation length, $\ell_{c}$ : for a turbulence spectrum of magnetic perturbations that follows a power law of index $\beta$ the following law, which is known in weak turbulence theory, has been extended into the regime of strong turbulence and large rigidities such that $r_{L} < \ell_{c}$ \citep{Casse01}:     
\begin{eqnarray}     
\bar{\ell}=\frac{r_{L}}{\eta_t}\,\left(\frac{r_{L}}{\ell_{c}}  \right)^{1-\beta}\,,     
\end{eqnarray}      
where $\eta_t =\frac{<\delta B^{2}>}{<B^2>}$.  The Bohm scaling $\bar{\ell} \sim r_{L}$, which holds for electrostatic turbulence, does not apply for purely magnetic irregularities on large scales; no theory nor numerical simulation has confirmed Bohm's conjecture. The Bohm estimate corresponds only to the specific case where the magnetic field is totally disorganised and the Larmor radius is as large as the correlation length, which is not the case in GRBs.\\     
According to a Kolmogorov scaling with $\beta =5/3$, we have shown \citep{GialisPelletier03} that GRBs are unable to produce UHECRs with this acceleration process because of a strong expansion limitation in energy      
\begin{eqnarray}     
     \epsilon_{exp} \simeq 10^{4}\left(\frac{\kappa_{0}}{10}\right)^{-3}\left(\frac{\eta}{300}\right)     
     \left(\frac{B_{\star}(r_{b})}{10^{4}G}\right)     
     \left(\frac{r}{r_{b}}\right)^{1-\alpha}\, GeV \,,     
     \label{eq:EEXP}     
\end{eqnarray}      
where $\kappa_0$ is the ratio of the acceleration time over the Larmor time for a Larmor radius that equals the correlation length of the magnetic field. This limitation, measured in the co-moving frame, is more severe than the synchrotron one and suggests that we have to consider another type of process to achieve high energy. This is the purpose of Sect. \ref{sec4}.    
    
\subsection{The magnetic field evolution} 
\label{sec23}     
   
The flow that reaches the beginning of the acceleration range is disconnected from the source if  
$c\Delta t_w < r_b$. In that case, the sheets that compose the flow start to expand with carrying a magnetic field that decays like $r^{-2}$, as argued by \cite{Meszaros93}. This is the most severe assumption on the evolution of the magnetic field in the attempt to accelerate particles. Some authors \citep{Waxman03} claimed that it decays like $r^{-1}$, because this would be consistent with the synchrotron interpretation of the GRB spectrum; this would require an amplification of the magnetic field in the vicinity of the shocks. We think that this would occur only in narrow layers behind the internal shocks and would be beneficial for electron acceleration. As for  the high energy protons, especially those candidates for UHECRs,  they necessarily explore a major part of the flow because of their large Larmor radii, and they experience a magnetic field that decays faster than $r^{-1}$, unless the flow is still connected with the source and carries a current associated with a toroidal field. In computations, for the sake of compromise, we have considered a field that decays like $r^{-3/2}$, which means that the magnetic energy is conserved in the shells. However, to demonstrate the capability of GRBs to produce UHECRs, we will also investigate the conservative case of a magnetic field decaying like $r^{-2}$.  
  
Because we will need some magnetic field intensity at the beginning of the acceleration range, we need to assume $B_0 \sim 10^{15}$ G at the black hole scale $r_0 \sim 10^7$ cm. However, for the work presented in this paper, we do not consider a situation where the magnetic field would dominate during the adiabatic expansion. The case of a magnetic field dominated expansion is interesting from the point of view of particle acceleration, but requires different modeling, postponed for future work.  
     
\section{The distribution function resulting from internal shocks}     
\label{sec3}     
     
In this paper, we will introduce an acceleration process additionnal to the usual internal shock acceleration. However we will consider this additionnal process as operating  above 1 GeV and below the expansion cut off mentioned in Sect. \ref{2.2} (see Eq. (\ref{eq:EEXP})). This yields an $\epsilon^{-2}$-spectrum in the frame $\mathcal{R}_\Gamma$ moving with a Lorentz factor $\Gamma$, in which the distribution function is almost isotropized. In fact, we will see later on that the highest energy part, especially that part accelerated by the additional process, is not fully isotropized. Because we will present numerical results in the observer frame (or, more properly, in the stationary frame of the source $\mathcal{R}_0$), it is useful to show how this truncated distribution looks in this frame.\\        
     
In the co-moving frame $\mathcal{R}_\Gamma$, the distribution is almost isotropized by turbulence, at least for the lower energy part, and the phase-space distribution function can be written      
\begin{eqnarray}     
f^{co}(\epsilon_{co},\mu_{co})= K\,\epsilon_{co}^{-4}     
\end{eqnarray}     
with $K = \frac{n_{\star}}{4\pi \ln \gamma_{m}}$ for $1<\epsilon_{co}<\gamma_{m}$ and it is null otherwise. The Lorentz factor $\gamma_{m}$ corresponds to the cut-off energy due to expansion.\\     
     
In the stationary frame $\mathcal{R}_0$, the distribution is highly anisotropic, since the proton energy, $\epsilon$, is such that $\epsilon = \delta(\mu)\,\epsilon_{co}$ where $\delta(\mu)\equiv 1/(\Gamma\,(1-\beta_0\,\mu))$.  The angular distribution is such that   
\begin{eqnarray}     
\rho_0(\mu) = \frac{1}{2\Gamma^{2}\,(1-\beta_{0}\,\mu)^{2}}\,,     
\label{eq4b}     
\end{eqnarray}     
where $\Gamma \gg 1$ implies values very close to 1 for $\mu$.\\     
   
The Lorentz invariance of the phase-space distribution function leads to an energy distribution function $\bar f$ such that     
\begin{eqnarray}     
\bar{f}(\epsilon)\equiv 2\pi\,\epsilon^{2}\int_{-1}^{1}f(\epsilon,\mu)\,d\mu= \frac{2\pi\,K\,\epsilon^{-2}}{\beta_0\,\Gamma}\,\int_{1/2\Gamma}^{2\Gamma}P(\delta)\,\delta^{2}\,d\delta    
\label{barf}  
\end{eqnarray}     
with $P(\delta) = 1$ for $\epsilon/\gamma_{m}<\delta<\epsilon$ and 0 otherwise.\\     
        
In the frame $\mathcal{R}_0$, one can summarize the effect of the anisotropy on the shape of the energy distribution function produced by the internal shocks as follows (see appendix) :     
\begin{eqnarray}     
\bar{f}(\epsilon)\propto \left\{      
\begin{tabular}{l}     
$\epsilon$ for $\epsilon <2\Gamma$\\      
$\epsilon^{-2}$ for $2\Gamma<\epsilon<2\Gamma\,\gamma_{m}$\\     
0 otherwise     
\end{tabular}\right..     
\end{eqnarray}     
     
As we will see in Sect. \ref{sec5}, a numerical simulation gives an illustration of the previous results about this distribution function resulting from a first stage of internal shock acceleration. It constitutes the "initial" condition of an additional Fermi acceleration process that we consider in the next section.     
     
\section{Relativistic Fermi acceleration by multiple fronts}     
\label{sec4}

In this section, we present the properties of hydromagnetic fronts in GRBs and we show that  
the Fermi acceleration process resulting from scattering off relativistic magnetized fronts  constitutes an interesting solution for UHECR generation.        
    
\subsection{Characterization of the hydromagnetic fronts in GRBs}     
\label{sub3.2}     
   
The internal shock model accounts for the light curves of GRBs that display fast variations, sometimes on a millisecond scale. This typically requires a number $N_s$ of sheets around 10-100, with Lorentz factors $\Gamma_k$ ($k=1,2,...,N_s$) between $10^2$-$10^3$ with respect to the frame  $\mathcal{R}_0$ \citep{Daigne98}. This means that they are mildly relativistic in the co-moving frame $\mathcal{R}_\Gamma$ with Lorentz factors $\gamma_k \simeq 2$, and we will refer to an average value $\gamma_\star$. For instance, assume that half of the sheets run forwards at a velocity $\beta_\star$ with respect to $\mathcal{R}_\Gamma$, and half of them run backwards at a velocity $-\beta_\star$, then we have a lower Lorentz factor $\Gamma^-_{\star} = \Gamma/2\gamma_{\star}$ and an upper Lorentz factor $\Gamma^+_{\star} = 2\gamma_{\star}\Gamma$ with respect to $\mathcal{R}_0$.   
   
The cosmic rays can be scattered by these sheets provided that they carry a sufficiently high magnetic field. Indeed they are scattered by magnetic sheets if and only if their Larmor radius is smaller than the thickness $\delta_k$ of the magnetic profile of the considered sheet. Particles of larger Larmor radius cross the front with just a small pitch angle jump of order $\delta_k/r_{L}$. We assume that $\delta_k$ is of the order of the thickness of the sheet and follows its evolution. Therefore, we assume an average thickness $\delta \simeq r/\Gamma^{2}$ measured in the stationary frame  $\mathcal{R}_0$. Consequently, particles of energy larger than $Z\,e\,\Gamma \,B_{\star}\,\delta$, where $B_{\star}$ is the average magnetic field in a sheet located at radius $r$, have a negligible interaction with the magnetic front. Thus, it is important to specify the interaction limit, $\epsilon_k$, for each sheet along the flow. We adopted the following scheme : we assume a prescribed value of the magnetic field $B_{\star}(r_{b})$ for each sheet at the distance $r_{b}$ and then any sheet carries a magnetic field of intensity decreasing with the distance $r$ like $r^{-\alpha}$ with $\alpha\in [1,2]$ (we will look at the case $\alpha = 3/2$ and $\alpha =2$ for the reason explained in Sect. \ref{sec23}). Now in the frame of each sheet, the particles are scattered with no  preferred direction and we can adopt either a mirror reflection or a random scattering. Both will give similar energy gains as will be seen later on; however we will choose the random scattering for the numerical simulation.\\     
Very high intensities of the magnetic field are allowed if we consider that the fireball results from the merging/collapse of a compact neutron star, namely $10^{15}$ G. For instance, a decrease of the magnetic field with $\alpha=3/2$ and an average Lorentz factor of 300, lead to $B_{\star}(r_{b})\simeq 3\times 10^{6}$ G for 100 sheets. We emphasize that this is a maximum value for a given number of sheets.  Thus the specification of the magnetic field is crucial, for it controls both the acceleration through scattering of particles having an energy lower than the interaction limit $\epsilon_k \propto r^{1-\alpha}$ and the spreading and decimation of the scattering sheets, responsible of both acceleration and expansion loss.\\    
   
Because of the forward and backward, mildly relativistic, motions of the sheets with respect to the co-moving frame $\mathcal{R}_\Gamma$, no isotropic distribution function can be built, only an axisymmetric and mirror-symmetric distribution is obtained. Assuming for instance a random scattering scheme (i.e. tendency of isotropization by each sheet), we obtain an angular distribution in this frame which is close to a superposition of two anisotropic distributions generated through the isotropic scattering by both forward (moving at $+\beta_{\star}$) and backward (moving at $-\beta_{\star}$) perturbations :  
\begin{equation}   
\label{ADF }   
\rho(\mu) = \frac{1}{2\gamma_{\star}^2} \frac{1+\beta_{\star}^2\mu^2}{(1-\beta_{\star}^2\mu^2)^2}\,. \    
\end{equation}    
     
\subsection{The relativistic Fermi process}     
     
The efficiency of the Fermi process with relativistic scattering fronts for the acceleration of supra-thermal particles was emphasized by \cite{Pelletier99}. This process consists in an interaction of a relativistic particle having an initial 4-vector momentum $(\epsilon_{1},\vec{p}_1)$ in the stationary frame $\mathcal{R}_0$ with a magnetized front  of velocity $\vec{\beta}_{k}$ (with corresponding $\Gamma_k$). We denote by primes physical quantities in the magnetized front frame $\mathcal{R}_k$ and we define the angle  between $\vec{\beta}_{k}$  and $\vec{p}_1$ by its cosine $\mu_{1}$ just before the interaction. The Lorentz transformation gives us the following equations :     
\begin{eqnarray}     
p_{1}' & \simeq & \Gamma_{k}\left( 1 - \beta_{k}\mu_{1} \right) p_{1}\,, \label{eq3}  \\     
\mu_{1}' & \simeq & \frac{\mu_{1} - \beta_{k}}{1 - \beta_{k} \mu_{1}}\label{eq4}\, ,      
\end{eqnarray}     
and assuming $\epsilon_{1}\simeq p_{1}\,c$ ($\geq 1$ GeV). \\     
  
The scattering does not change the particle energy in the magnetized front frame and the pitch angle (cosine) $\mu_{1}'$ is randomly changed into $\mu_{2}'$.       
Denoting by 2 the physical quantities after the interaction, the reversed Lorentz transformation leads to    
\begin{eqnarray}     
p_{2} & \simeq & \Gamma_{k}^{2}\left( 1 + \beta_{k}\mu_{2}' \right)\left( 1 - \beta_{k}\mu_{1} \right) p_{1}  \label{eq5}\,,\\     
\mu_{2} & \simeq & \frac{\mu_{2}' + \beta_{k}}{1 + \beta_{k} \mu_{2}'} \label{eq6}\,.       
\end{eqnarray}    
However to estimate the energy gain, it is necessary to calculate this in the frame where the distribution function is even in terms of the cosine of the pitch angle, namely in $\mathcal{R}_\Gamma$.   
The following unprimed quantities are now measured in $\mathcal{R}_\Gamma$. Therefore the energy gain per scattering spreads from $(1-\tilde\beta_k)/(1+\tilde\beta_k) \simeq 1/4\gamma_{k}^2$ (for rear-on collision) to $(1+\tilde\beta_k)/(1-\tilde\beta_k) \simeq 4\gamma_{k}^2$ (for head-on collision), the $\tilde\beta$ being the velocities measured in $\mathcal{R}_\Gamma$. Since the distribution function is even in $\mu$ in $\mathcal{R}_\Gamma$, the average energy gain per interaction is  $\simeq \gamma_{\star}^2$ for both the random scattering and the mirror reflection schemes, $\gamma_{\star}$ being the average value of the $\gamma_k$. For example, a proton which initially has an energy of 100 GeV could achieve an energy of $10^{10}$ GeV after about 12  interactions with $\gamma_{\star}= 2 $, whereas 6 Fermi cycles would amplify its energy by a factor 64 since the gain per Fermi cycle remains smaller than 2 (after the first one) \citep{Gallant1999}. Therefore the interest of multi-front interactions compared to the acceleration in a single relativistic shock is that the full energy gain $\sim \gamma_{\star}^{2}$ can be obtained at each scattering with crossing fronts, and could generate UHECRs if the cosmic rays undergo enough scatterings off the magnetized fronts. We will see further on that this required number of scatterings puts a constraint on the number of sheets.

\section{Description of the numerical simulation}     
\label{sec5}     
     
\subsection{Initial conditions}     
     
\begin{figure*}[h!t]     
\begin{center}     
\includegraphics[totalheight=6.4cm,angle=-90]{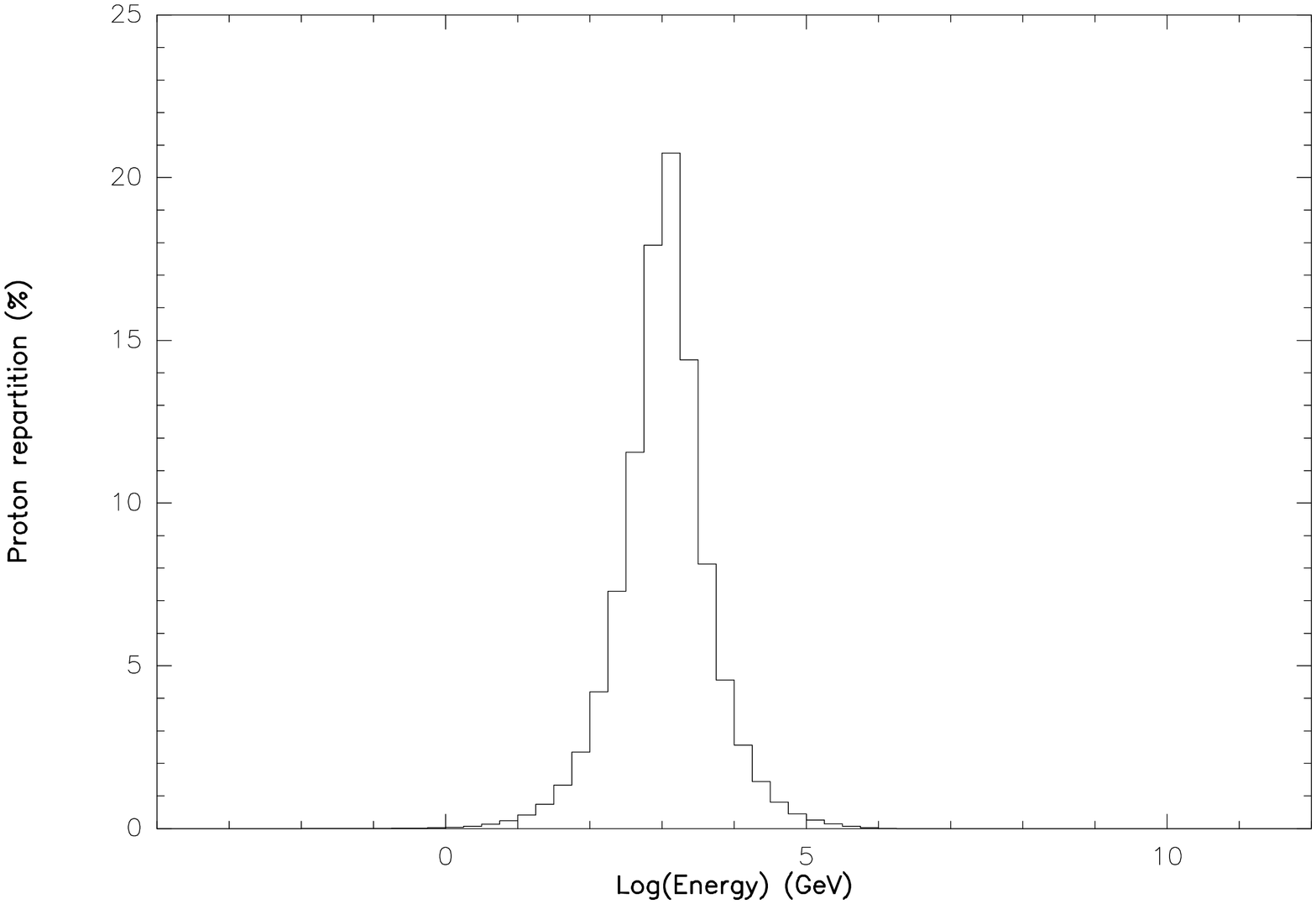}     
\includegraphics[totalheight=6.4cm,angle=-90]{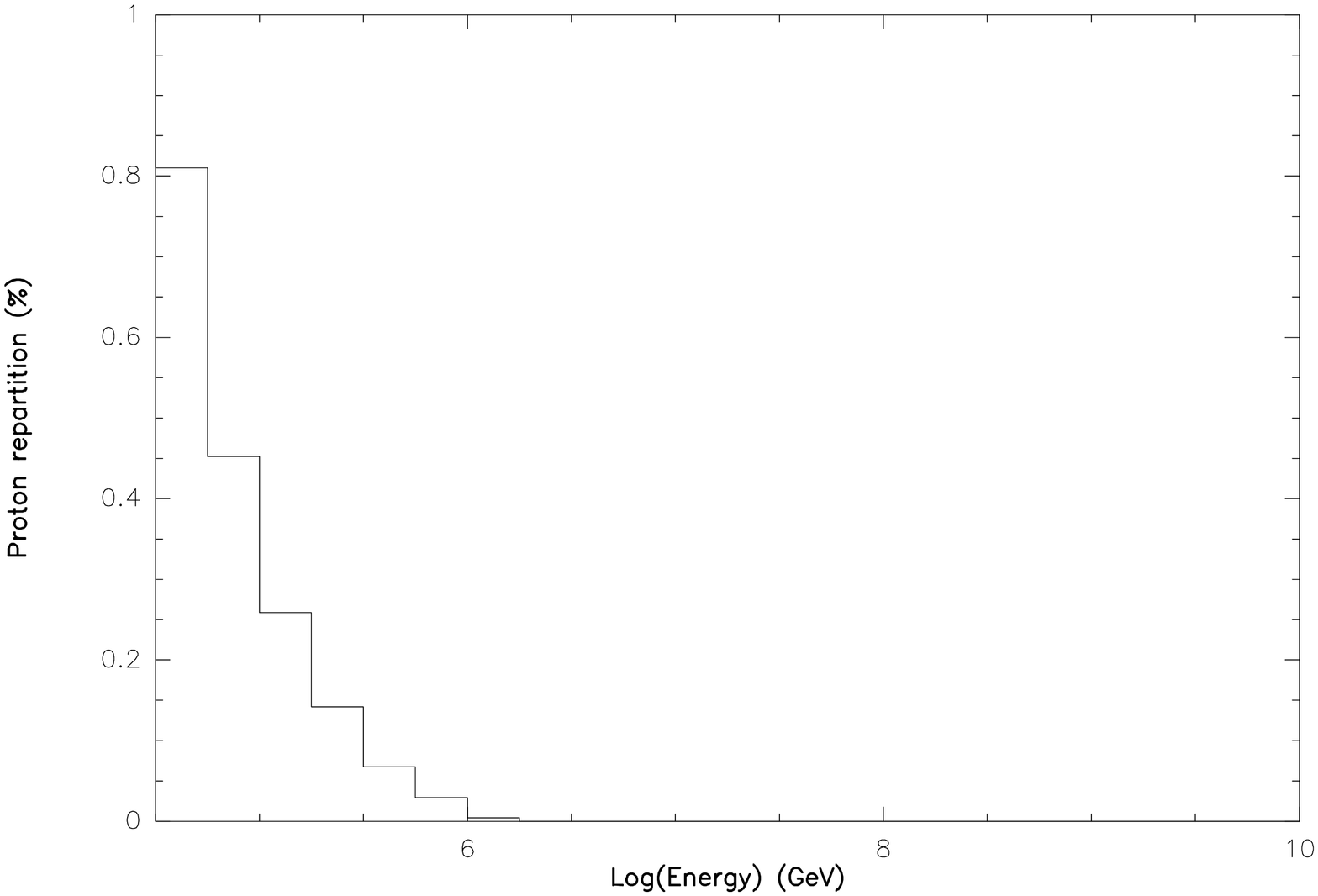}     
\includegraphics[totalheight=6.4cm,angle=-90]{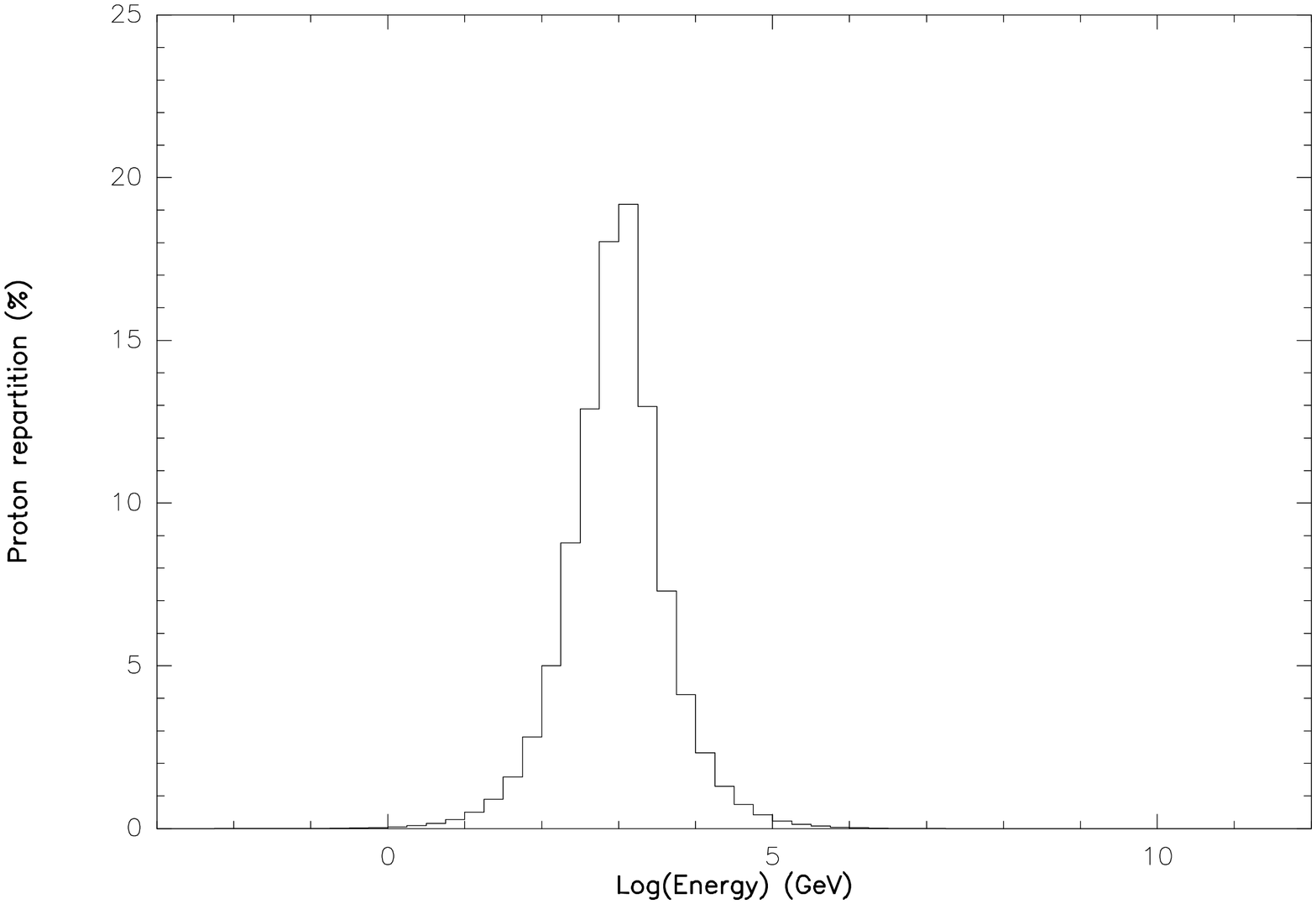}     
\includegraphics[totalheight=6.4cm,angle=-90]{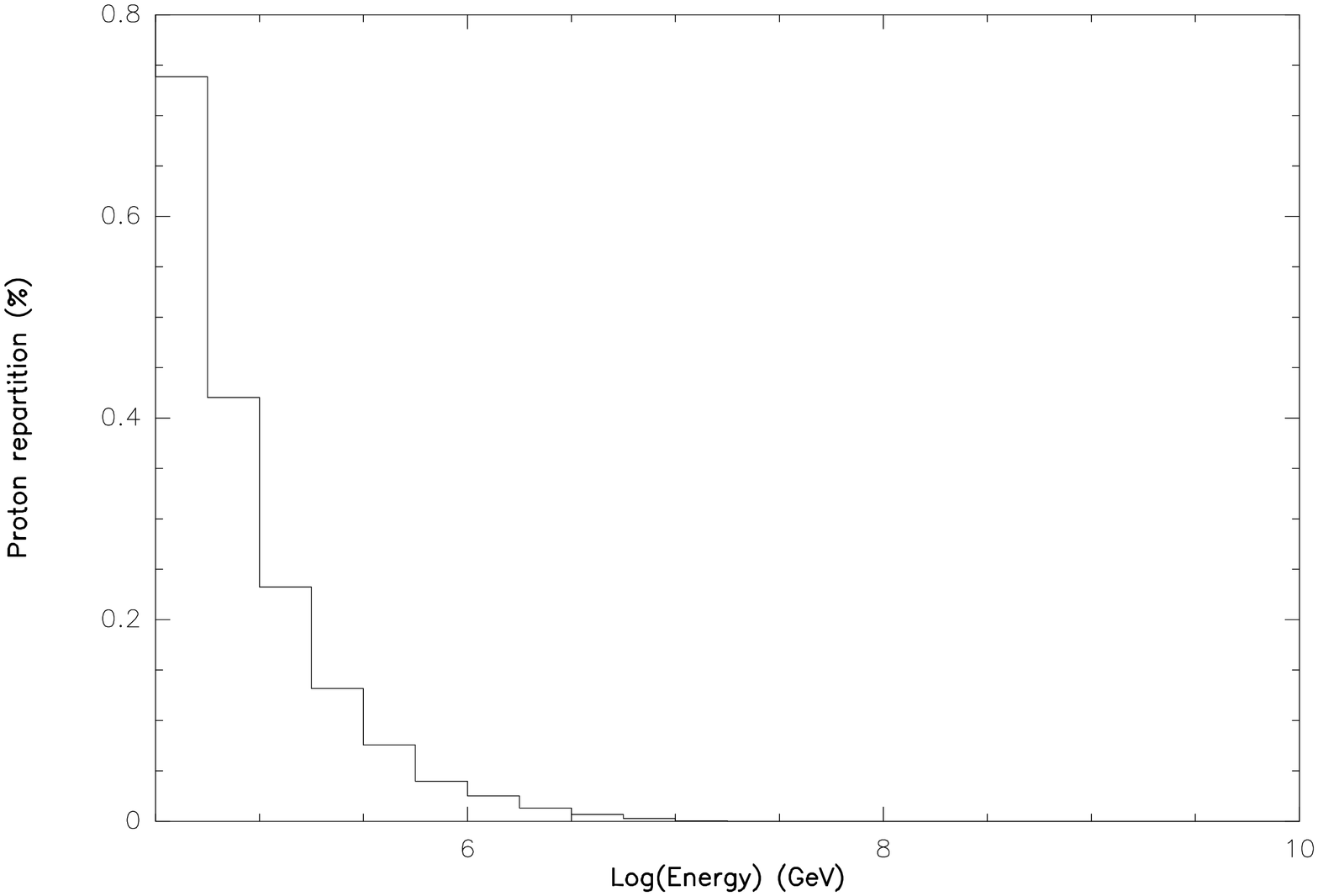}     
\caption{The proton repartition in the stationary frame $\mathcal{R}_{0}$ calculated with the simulation and before the additionnal acceleration Fermi process : this proton population is just the result of a statistical superposition of the contribution of every sheet. Top for a magnetic field at $r_{b}$ of $10^{5}$ G. Bottom: $B_{\star}(r_{b})=10^{6}$ G. According to the initial distribution function calculated in Sect. \ref{sec3} and the cut-off energy (Eq.(\ref{ec})), the higher the magnetic field, the more the distribution spreads over high energy.}     
\label{fig1}     
\end{center}     
\end{figure*}

We have considered a conical box representing the ultrarelativistic flow around the broadening radius $r_{b}$ ($\simeq 10^{12}$ cm). In the stationary frame $\mathcal{R}_{0}$, the initial box height, corresponding to the flow length, is given by $c\,\Delta t_{w}$ and the transversal size of the flow is equal to $\sqrt{\Omega/\pi}\,r$ with $\Omega \simeq 4\pi/500$, the opening angle of the flow.      
Using a Monte-Carlo simulation, we randomly put in the flow a set of identical (same mass) baryonic layers (or sheets) with an uniform Lorentz factor distribution between $10^{2}$ and $10^{3}$ considered by several authors \citep{Daigne98, Piran00} to reproduce the GRB light curves from internal shocks. Following previous simulations \citep{Kobayashi97}, we consider a number of layers of about several times ten. For a layer with a Lorentz factor of $\Gamma_{k}$ at a distance $r$ the width is equal to $r/\Gamma_{k}^{2}$ and the transverse radius is the same as that of the flow.\\     
The isotropic suprathermal proton population is initially injected in each layer with an energy spectrum in the co-moving frame proportional to ${\epsilon'}^{-2}$ between 1 GeV and a cut-off energy      
\begin{eqnarray}     
\epsilon_{c}=10^{4}\,\left( \frac{B_{\star}(r_{b})}{10^{6}\, G}\right) \,\,GeV\,,     
\label{ec}     
\end{eqnarray}       
according to the strong expansion limitation in energy (see Eq. (\ref{eq:EEXP})).   
In the stationary frame $\mathcal{R}_{0}$, a Lorentz transformation provides the energy, $\epsilon$, of a proton      
\begin{eqnarray}     
\epsilon = \Gamma_{k}\,(1\pm \beta_{k}\,\beta^{c}_{r})\,\epsilon' \,,     
\end{eqnarray}     
where $\beta^{c}_{r}$ is the radial velocity of the proton in the co-moving frame of the sheet and $\epsilon'$ its energy.  One can note that the energy $\epsilon'$ of a proton in the co-moving frame $\mathcal{R}_{k}$ can be amplified by a factor 2$\Gamma_{k}$ in the stationary frame.\\     
For each proton, we initially define a random pitch angle cosine as defined in Sect. \ref{sec3} (see Eq. (\ref{eq4b})). We have plotted in Fig. \ref{fig1}, for $B_{\star}(r_{b})=10^{5}$ G and  $B_{\star}(r_{b})=10^{6}$ G, the initial energy repartition of the proton population, in the observer frame, which depends on three independent random parameters namely $\Gamma_{k}$, $\beta^{c}_{r}$ and $\epsilon'$.     

\subsection{Temporal evolution}     
     
We describe here the temporal evolution in our Monte-Carlo simulation. The Lorentz factor distribution leads to collisions of different sheets during the numerical process between $r_{b}$ and $300\,r_{b}$ (about $r_{d}/3$ for $r_{d}=10^{15}$ cm or $r_{d}/30$ for $r_{d}=10^{16}$ cm) which is sufficient to determine the fraction of UHECRs according to the confinement limit (see Sect. \ref{sec2}). We have considered the simple case where the result of a collision of two identical sheets is a single sheet having a Lorentz factor defined by $\sqrt{\Gamma_{1}\,\Gamma_{2}}$ where $\Gamma_{1}$ and $\Gamma_{2}$ are the Lorentz factors of the sheets before colliding. The total number of sheets decreases with time and this reduces the interaction probability with protons in the flow even if the sheet width and the transversal radius increase like $r$ in our model. \\     
The first collision of a sheet defines the time from which its protons are released into the flow and can interact with all sheets. After a proton-sheet interaction, which is spatially defined by comparing respective positions, we randomly assign a new pitch angle for the proton and its energy is changed according to the equation (\ref{eq5}). We assume that all interactions are instantaneous. Indeed, the magnetic scattering duration inside a sheet, $\tau_{s}$,  can be neglected; for a proton with energy $\epsilon'$ in the co-moving frame, it can be expressed by       
\begin{eqnarray}     
\tau_{s}=2.3 \gamma_{\star}\times 10^{-12} \left(\frac{B_{\star}}{10^{6}\,G}\right)^{-1}\left(\frac{\epsilon'}{1\,GeV}\right)\,s\,,     
\end{eqnarray}       
and it is very small compared to the mean free path time $r/cN_s > 1$ s.  \\     
The proton energy limitation for an interaction (or magnetic scattering) is given by the interaction limit calculated in each layer (see Sect. \ref{sec2} and paragraph \ref{sub3.2}); for a given proton, the higher its energy is, the more transparent the medium is to it and its escape probability becomes important. This limitation is far more important than the transverse escape.  
 We also have neglected the synchrotron losses for protons because the synchrotron limitation, in the co-moving frame $\mathcal{R}_{\Gamma}$, is over $10^5$ GeV at $r_{b}$ and next, increases like $r^2$ for $B_{\star}\propto r^{-2}$ and $r^{11/8}$ for $B_{\star}\propto r^{-3/2}$ \citep{GialisPelletier03}.

\section{Numerical results and discussion}     
\label{sec6}

\subsection{Generation of UHECRs}

\begin{figure*}[h!t]     
\begin{center}     
\includegraphics[totalheight=6.4cm,angle=-90]{0301fig2.ps}     
\includegraphics[totalheight=6.4cm,angle=-90]{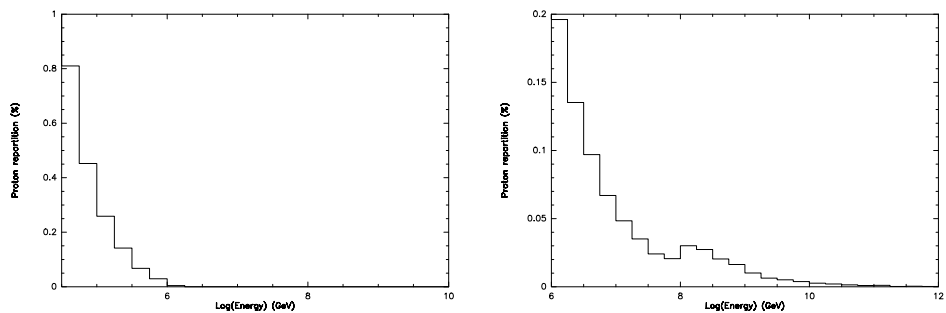}     
\caption{The initial proton distribution (left figure) is strongly modified by the Fermi acceleration process in the high energy range where an UHECR population appears (right figure) with $B_{\star}(r_{b})=10^5$ G, $\Delta t_{w}=10$ s, $N_{s}=50$.}     
\label{fig2}     
\end{center}     
\end{figure*}

The numerical Monte-Carlo simulation clearly shows that a sizeable part of cosmic rays is produced by the relativistic Fermi process stretching the initial distribution tail up into the UHE range (see Fig. \ref{fig2}); for any magnetic field intensity higher than $10^{4}$ G at $r_{b}$, the relativistic Fermi acceleration by multiple fronts leads to a production of UHECRs with a fraction larger than $10^{-6}$ for a given initial proton distribution as prescribed in Sect. \ref{sec5}. Table \ref{restab} summarizes the results for different values of $B_{\star}(r_{b})$ and of the flow duration. We have chosen a typical number of sheets, $N_{s}$, of 50 at the beginning of the process and, first, a magnetic field that decreases like $r^{-3/2}$, second, a magnetic field that decreases like $r^{-2}$.     
     
\begin{table*}[h!t]     
\begin{center}     
\begin{tabular}{|c|c|c|c|}      
\hline     
Flow duration (in s)     
& \multicolumn{3}{c|}{Magnetic field at $r_{b}$}\\     
& $10^{4}$G & $10^{5}$G & $10^{6}$G \\      
\hline     
0.5 & $5.4\times 10^{-5}$ & $1.5\times 10^{-4}$  & $3.6\times 10^{-4}$ \\     
\hline     
1.0 & $6.3\times 10^{-5}$ & $1.9\times 10^{-4}$ & $4.1\times 10^{-4}$\\     
\hline     
5.0 & $5.1\times 10^{-5}$ & $1.8\times 10^{-4}$ & $4.8\times 10^{-4}$\\     
\hline     
10.0 & $2.0\times 10^{-5}$ & $1.2\times 10^{-4}$ & $3.2\times 10^{-4}$\\     
\hline     
\end{tabular}     
\caption{The UHECR ($>10^{10}$ GeV) fraction generated by relativistic Fermi acceleration in the Monte-Carlo simulation. A number of 50 layers is considered here and the flow duration varies from 0.5 s (short GRB) to 10.0 s (long GRB). The magnetic field decreases as $r^{-3/2}$.}    
\label{restab}     
\end{center}      
\end{table*}

\begin{figure*}[h!t]     
\begin{center}     
\includegraphics[totalheight=7.5cm,angle=-90]{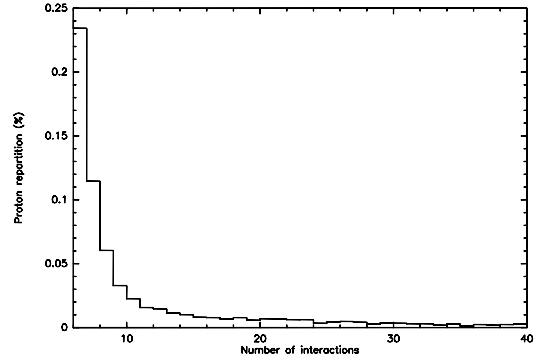}     
\caption{The proton repartition as a function of the number of interactions (or scatterings) for a magnetic field at $r_{b}$ of $10^{4}$ G and for a wind duration $\Delta t_{w}=10$ s: a sizeable fraction of protons ($> 2.0\times 10^{-3}$) undergoes a sufficient number of interactions ($\geq 10$) and this makes the generation of UHECRs possible.}     
\label{fig4}     
\end{center}     
\end{figure*}

\begin{figure*}[h!t]     
\begin{center}     
\includegraphics[totalheight=6.4cm,angle=-90]{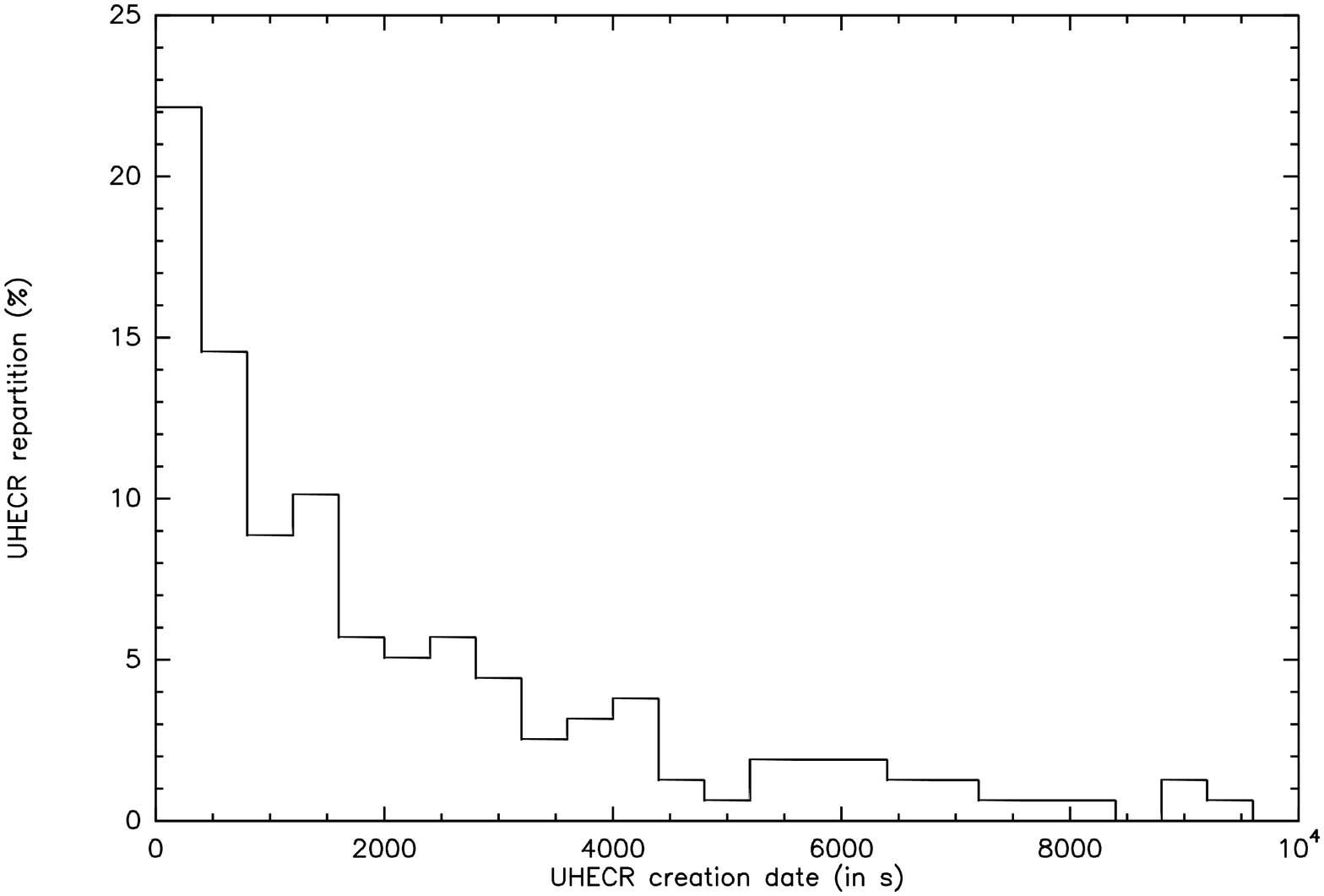}     
\includegraphics[totalheight=6.4cm,angle=-90]{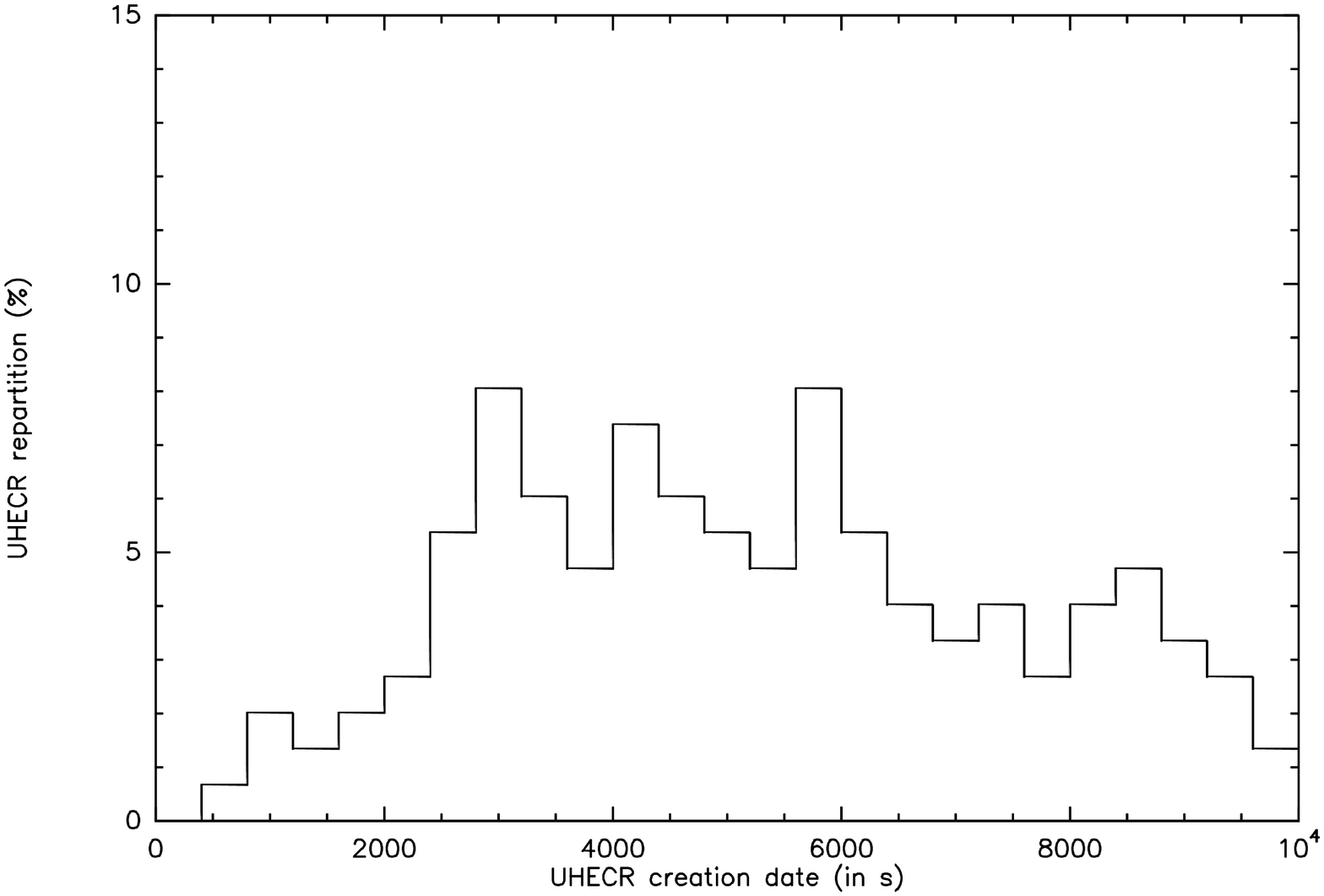}     
\includegraphics[totalheight=6.4cm,angle=-90]{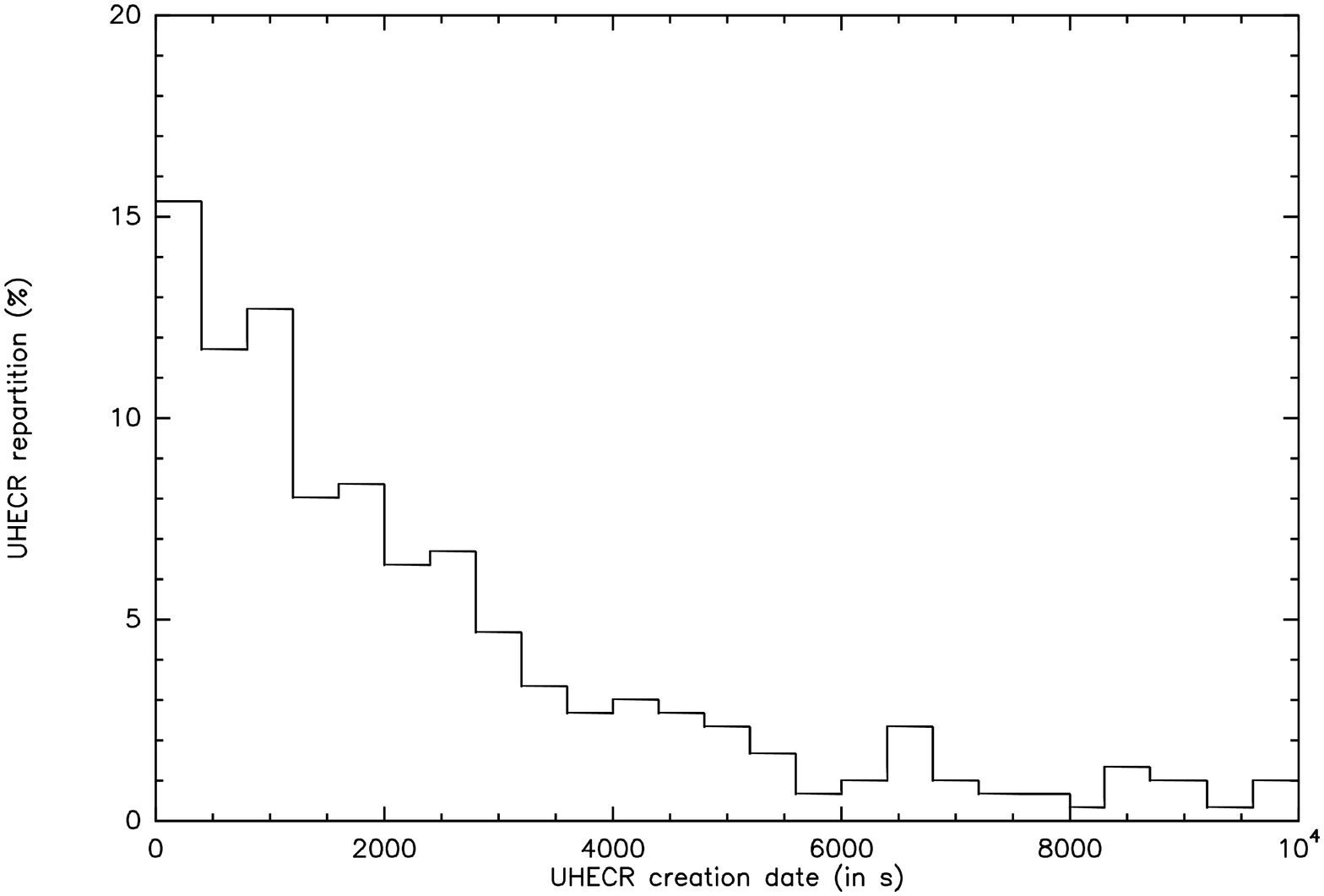}     
\includegraphics[totalheight=6.4cm,angle=-90]{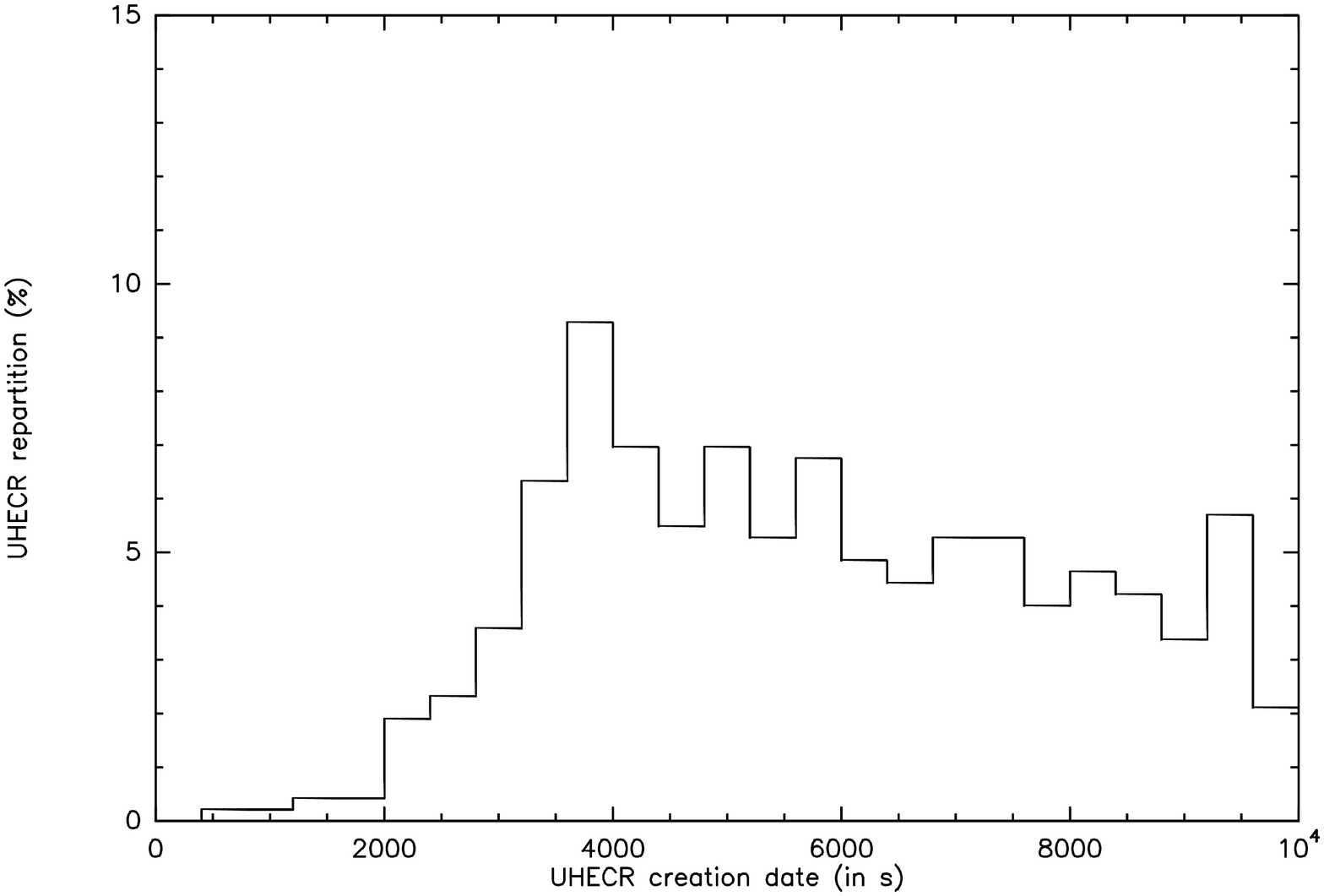}     
\caption{The UHECR repartition (in $\%$ of the total UHECR population) as a function of the UHECR creation time which mainly happens between $r_{b}$ ($t=0$ on the figures) and about $300\,r_{b}$ ($t=10^{4}$ s) in the Monte-Carlo simulation. Top a magnetic field at $r_{b}$ of $10^{4}$ G : (left $\Delta t_{w}=0.5$ s, right $\Delta t_{w}=10$ s). Bottom magnetic field at $r_{b}$ of $10^{5}$ G (left $\Delta t_{w}=0.5$ s and right $\Delta t_{w}=10$ s). }     
\label{fig3}     
\end{center}     
\end{figure*}

The determination of the time of UHECR creation which has been made for different values of the wind duration and the magnetic field (see Fig. \ref{fig3}) shows that UHECR generation mainly happens at the very beginning of the internal shock phase; more precisely, UHECR generation can be achieved before $\sim 300\,r_{b}$ (or $\sim 10^{4}$ s) which corresponds to only a few percent of the internal shock phase duration ($\gtrsim 10^{5}$ s). There are two conditions that govern the ending of the multi-front acceleration: first, the maximum value of the interaction limit $\epsilon_k$ assigned at the radius $r_b$ through the value of the magnetic field there; and second, the decrease of the interaction energy limit with distance. Indeed, the lowest interaction energy limit (see Sect. \ref{sec2}) to get UHECRs is about $10^{19}$ eV at $r_{b}$ for $B_{\star}(r_{b})=10^{4}$ G; the difference between this energy and the UHE range ($\gtrsim 10^{20}$ eV) is almost at the maximum possible value of the energy jump (the maximum gain being about 16), and then the gap quickly becomes too high with time to be jumped by a proton in a single proton-sheet interaction, since the interaction limit decreases like $\epsilon_k \propto  r^{1-\alpha}$. However, we have shown (see Sect, \ref{sec4}) that only a small number of interactions ($\gtrsim 10$) is necessary to achieve the UHE range. The simulation directly provides this number per proton: whatever the flow duration between 0.5 and 10 s and for a magnetic field of $10^{4}$ G at $r_{b}$, the results confirm that a sizeable fraction of protons ($> 2.0\times 10^{-3}$) undergoes more than 10 interactions with the hydromagnetic fronts and, consequently, can easily achieve the UHE range (see Fig. \ref{fig4}).    
Moreover, Table \ref{restab} indicates that the energy amount converted into UHECRs at the beginning of the Fermi process is high enough to constitute a sizeable fraction of the total magnetic energy. \\    
 Finally, a magnetic field decreasing in $r^{-2}$ leads to a lower resulting fraction of UHECRs but a sizeable one again: for instance, for a magnetic field $B_{\star}(r_{b})=10^4$ G, this fraction is about $8.0\times 10^{-6}$ with $\Delta t_{w}=10$ s and $2.6\times 10^{-5}$ with $\Delta t_{w}=0.5$ s.\\    
Let us now give arguments concerning the  number of sheets required to produce UHECRs. The numerical simulation shows that the acceleration stops because of the decimation of the sheets due to expansion (the mean free path increases like $\bar \ell_s \propto r$) and mostly due to decrease of the interaction energy limits ($\epsilon_k \propto r^{-1}$ for $B_{\star} \propto r^{-2}$) which rarefies the scattering medium for increasing energies. Since $\bar \ell_s/r$ is given  by the number of sheets $N_s$, namely $\bar \ell_s / r = 1 / N_s$, a simple estimate of the minimum number of sheets can be given. At each scattering, a particle migrates towards a region of lower interaction limit and roughly, after $n$ scatterings $\epsilon_{s} (n) = (1-q)^n \epsilon_s(0)$, where $q= (\alpha -1)/N_s$ and $\epsilon_s$ denotes the local scattering energy limit, rather than the interaction limit associated to a specific sheet. So the acceleration stops when the number of scatterings is such that    
$\gamma_{\star}^{2n} \epsilon_0 \sim (1-q)^n \epsilon_s(0)$, which leads to the following number of scatterings:   
\begin{equation}   
\label{ }   
n_c \sim \frac{\ln \left( {\epsilon_s(0)\over \epsilon_0}\right) }{\ln \left({\gamma_{\star}^2 \over 1-q}\right)}   
\end{equation}   
Since we expect that the energy reached after $n_c$ scatterings is not far from the maximum value of the interaction limit, it requires $n_c\, q <1$, and thus    
\begin{equation}   
\label{ }   
N_s > \frac{\ln \left( \epsilon_{s}(r_b)\over \epsilon_0\right) }{\ln \gamma_{\star}^2} \ .   
\end{equation}   
To get particles that will reach an energy at least equal to the interaction limit at $r_b$ with $\gamma_{\star} = 2$ and $B_{\star}(r_{b})=10^4$ G, we need more than 13 to 17 sheets.   

Since the process tends to produce cosmic rays that reach an energy close to the local scattering limit, the integrated distribution function reflects the evolution of the scattering limit, which is directly given by the magnetic field decay. For an intensity of the magnetic field decreasing like $r^{-2}$, the scattering limit decreases like $r^{-1}$, and for a continuous flow during $\Delta t_w$, the cosmic ray density decreases like $r^{-2}$. Therefore the integrated energy distribution function is  
\begin{equation}  
\label{IDF }  
{\ud N_{\star} \over \ud \epsilon} = \int \delta(\epsilon - \epsilon_s(r))\,n_{\star} \,\Omega\, r^2\,\ud r \propto \int \delta(\epsilon - \epsilon_s(r)) \,\frac{\ud\epsilon_s}{\epsilon_s^2} \propto \epsilon^{-2} \ .  
\end{equation}   
Thus we obtain an $\epsilon^{-2}$-spectrum over 4 decades, because the range of scattering limit energy stems from the acceleration range that extends from $r_b$ to $\sim r_d$.

\subsection{Diagnosis of Cosmic Rays in GRBs}     
     
\subsubsection{The two types of neutrino emission}     
     
There are two possible types of neutrino emission due to the generation of cosmic rays \citep{GialisPelletier03}. The most expected one results from the $p\gamma$-process where the total energy radiated by neutrinos in the co-moving flow is given by      
\begin{equation}     
     E'_{\nu} \sim     
     \alpha_{0}\,\xi_{\nu}N_{*}\log\left(\frac{\gamma_{max}}{\gamma_{th}}\right)\,\bar \epsilon'_{\gamma} \ ,     
     \label{eq:EPNU}     
\end{equation}     
where the threshold is at $\gamma_{th} \sim 10^4$. 
It can be easily established that even this improved acceleration process does not provide a sufficient increase of the neutrino emission. Taking account of the collimation, the number of $p\gamma$-neutrinos that cross a km$^2$-detector, coming from a GRB at $100 Mpc$, is about $10^4$.\\     
     
The other possibility is low energy emission between 5 GeV and 150 GeV resulting from $pp$-collisions. We predicted significant emission due to the standard Fermi acceleration at internal shocks when $\eta$ is smaller than $\eta_{\star}/2$ where $\eta_{\star}$ is reached when the photospheric radius is equal to $r_{b}$ \citep{GialisPelletier03}. The improvement of the acceleration by the process presented in this paper does not significantly change the order of magnitude of our previous estimate.  The number of $pp$-neutrinos  is greater and could allow the formation of a spectrum, since we could get $10^8$ neutrinos under the same conditions.

\subsubsection{Gamma emission by $\pi^{0}$-decay}     
     
The $pp$-process generates $\pi^{0}$-mesons that decay into two $\gamma$-photons. This emission deserves more attention because of a possible detection in the GeV range, which would constitute an interesting diagnosis. For a $\epsilon^{-2}$ spectrum of cosmic rays, the gamma flux due to $\pi^{0}$-decay is estimated by     
\begin{eqnarray}     
S_{\pi^{0}}(\epsilon_{\gamma})\sim c\,\sigma_{pp}\,n_{p}\,N_{\star}\,\left(\frac{\epsilon_{\gamma}}{\epsilon_{0}}\right)^{-2}\,,      
\label{}     
\end{eqnarray}     
for $\epsilon_{\gamma}>\epsilon_{0}$ and $S_{\pi^{0}}(\epsilon_{\gamma})=0$ otherwise, where $\epsilon_{0}\simeq 67$ MeV in the co-moving frame, which corresponds to about 20 GeV for the observer. Actually the fireball is optically thin to Compton scattering at 67 MeV and beyond, thanks to the Klein-Nishina regime, and also to gamma-gamma pair production for two possible target photons (as this will be detailed in a forthcoming paper).
This spectrum thus has a power law tail that extends over several decades; the highest part would reveal the presence of UHECRs, but would require a very high sensitivity. Detection of gamma-rays in that energy range from GRBs seems to be expected by ground gamma-ray telescopes, as suggested by the analysis of the spectrum of GRB970417a by the Milagro collaboration \citep{Atkins2000}.\\     
The $\pi^0$-decay spectrum must be compared with the SSC-spectrum. The latter can be written;     
\begin{eqnarray}      
S_{SSC}(\epsilon_{\gamma})\sim c\, \sigma_{T}\,n_{s}\,N^{\star}_{e}\,\left(\frac{\epsilon_{\gamma}}{\epsilon_{peak}}\right)^{-1/2}\,,     
\end{eqnarray}     
for $\epsilon_{\gamma}> \epsilon_{SSC}$ with $\epsilon_{SSC} \equiv \epsilon_{peak}\,\gamma^{2}_{peak}\sim 10^{14}$ eV (for $\epsilon_{peak}=1$ MeV for the observer and $B_{\star}(r_{b})=10^{4}$ G) the energy of the SSC-spectrum is maximum and $S_{SSC}(\epsilon_{\gamma})=0$ otherwise, where $N^{\star}_{e}$ is the relativistic fraction of the electron population.\\     
The diagnosis of the $\pi^{0}$-decay should be fairly easy since it radiates at much higher energy than the synchrotron emission and at much lower energy than the SSC-emission. When $\eta < \eta_{\star}/2$, the fraction of the energy radiated by $\pi^{0}$-decay is quite significant, a few percent; indeed     
\begin{eqnarray}     
\frac{E_{\pi^{0}}}{E}\sim 0.07\times \frac{\Delta t_{w}}{\eta^{2}\,t_{0}}\sim 10^{-2}-10^{-1} \,,     
\end{eqnarray}     
with $t_{0}\equiv r_{0}/c \sim 1$ ms.\\     
     
Such a detection (by HESS observatory, 5@5 experiment and GLAST) would provide a very nice estimate of the baryon load together with evidence of cosmic ray generation.

\section{Conclusion}     
     
We have shown through accurate investigation of scattering properties that the usual Fermi processes are unable to generate UHECRs in GRBs. First of all, it is well known that the highly relativistic external shock with a standard interstellar medium cannot achieve the goal of UHECR generation. The hope that it could do so was mostly based on the internal shock model but the irrelevant Bohm's scaling was used; moreover, a slowly decreasing magnetic field was assumed. Removing these extreme assumptions made the goal impossible \citep{GialisPelletier03}. \\     
However, because the global energy budget is in favour of copious cosmic ray generation, we thought it interesting to design an additional Fermi process that would stretch the cosmic ray distribution tail. Indeed, the acceleration through multi-scattering off magnetized relativistic fronts turns out to be more efficient even with a field decreasing in $r^{-2}$. In this paper, we found that a sizeable fraction ($\sim 10^{-5}$) of the cosmic ray population reaches the UHE range, which is even more than necessary. \\     
There is another important physical issue, namely the escape of the UHECRs from the fireball towards the observer. The relativistic Fermi process that we proposed generates UHECRs in a thin layer located at the very beginning of the internal shock phase. The question arises whether these particles suffer a further expansion loss. In fact, because the magnetic field decreases faster than $r^{-1}$, the interaction energy limit with the magnetic fronts decreases and the UHECRs are no longer scattered by them. Thus, because of the lack of scattering, they cannot experience the expansion loss, contrarily to models expecting $B\propto r^{-1}$. Therefore, they directly travel across the shells, benefiting of the Lorentz boost. For a magnetic field decreasing as $r^{-2}$, the integrated spectrum is in $\epsilon^{-2}$ over 4 decades.\\       
In our previous paper \citep{GialisPelletier03}, we predicted a possible double neutrino emission;  even though this improvement of the acceleration process does not significantly change the estimated fluxes, the collimation allows detectable events, especially with pp-neutrinos. However, we found a possible diagnosis of cosmic ray generation in some numerous class of GRBs, having $\eta < \eta_{\star}/2$, associated with the signature of $\pi^{0}$-decay in the GeV $\gamma$-ray spectrum. This seems to be really observable, with a few percent of the GRB energy, for GRBs having a sufficient baryon load. This signature  (in the GeV range by HESS observatory, 5@5 experiment and GLAST) would be easily distinguished from the synchrotron bump and the SSC-bump with a flux as high as a few percent of the fireball global energy flux.\\     
This study gives a new chance to GRBs as sources of UHECRs. But their contribution to the diffuse background spectrum might not be dominant compared to the AGN contribution. This is an important issue that is under intense debate in the literature \citep{Stecker2000,Berezinsky03,BahcallWaxman03}.

\section*{Appendix}   
     
From equation \ref{barf}, we deduce the following expression     
\begin{eqnarray}     
\bar{f}(\epsilon)=A\, \epsilon^{-2}\,(\delta_{2}^{3}-\delta_{1}^{3})\,,     
\end{eqnarray}     
where $A = n_{\star}/(6\ln \gamma_{m}+\Gamma\,\beta_0)$ and $\delta_{1}=\rm{Sup}\{\epsilon/\gamma_{m},1/2\Gamma\}$, $\delta_{2}=\rm{Inf}\{\epsilon,2\Gamma\}$.\\     
According to the equation (\ref{ecl1}) and for a typical magnetic field of $10^{5}$ G at $r_{b}$, we have supposed that $\gamma_{m}>2\Gamma$. It is noteworthy that the following calculation leads to the same result for the hypothesis $\gamma_{m}<2\Gamma$. Let us distinguish two cases : the first one for $\epsilon > \gamma_{m}$ gives $\delta_{1}=\epsilon/\gamma_{m}$ and $\delta_{2}=2\Gamma$. The distribution function becomes     
\begin{eqnarray}     
\bar{f}_{\epsilon > \gamma_{m}}(\epsilon)=A\, \epsilon^{-2} \left( 8\Gamma^{3}-\frac{\epsilon^{3}}{\gamma_{m}^{3}}\right)\propto\epsilon^{-2}\, ,     
\end{eqnarray}     
for $\epsilon < 2\Gamma\,\gamma_{m}$ and $\bar{f}_{\epsilon > \gamma_{m}}(\epsilon)= 0$ beyond $2\Gamma\,\gamma_{m}$.\\     
     
The second case is considered for $\epsilon < \gamma_{m}$ and it can be divided in two sub cases : first, for $\epsilon > 2\Gamma$, the distribution function can be simplified and becomes such that     
\begin{eqnarray}     
\bar{f}_{2\Gamma<\epsilon < \gamma_{m}}(\epsilon)\propto \epsilon^{-2}\,.      
\end{eqnarray}      
Finally, for $\epsilon<2\Gamma$, we have $\delta_{2}=\epsilon \gg \delta_{1}$ and the distribution function can be written     
\begin{eqnarray}     
\bar{f}_{\epsilon <2\Gamma} (\epsilon)\propto \epsilon\,.     
\end{eqnarray}

\bibliographystyle{aa}
\bibliography{/gagax1/ur1/dgialis/THESE/BIBFILES/biblio}

\end{document}